\documentstyle[12pt,aasms4, flushrt]{article}   
 
\def\lta{\la}
\def\gta{\ga}

\begin{document}

\title{ DOES THE THERMAL DISC INSTABILITY OPERATE 
 IN ACTIVE GALACTIC NUCLEI? }

\author{L. ~Burderi \altaffilmark{1,2}, A. ~R. ~King \altaffilmark{2} and
E. ~Szuszkiewicz \altaffilmark{2}}
\altaffiltext{1}{Istituto di Fisica dell'Universita', via Archirafi 36,
90123 Palermo, Italy}
\altaffiltext{2}{Astronomy Group, University of Leicester, Leicester
LE1 7RH, U.K.}
 
\authoremail{lbu@star.le.ac.uk, ark@star.le.ac.uk, esz@star.le.ac.uk}

\begin{abstract}
We examine all possible stationary, optically thick, geometrically
thin accretion disc models relevant for
active galactic nuclei (AGN) and identify the physical regimes in which
they are stable against the thermal--viscous hydrogen ionization instability.
Self--gravity and irradiation effects are included.
We find that most if not all AGN discs are unstable. 
Observed AGN therefore represent the
outburst state, although some or all quasars could constitute a steady
population having markedly higher fuelling rates than other AGN. 
It has important  implications for the AGN mass supply and for
the presence of supermassive black holes in nearby spirals.
\end{abstract}
 

\section{INTRODUCTION}

Accretion discs are a nearly ubiquitous feature of close binary
systems, and their presence is widely invoked in models of active
galactic nuclei (AGN). A major feature of the discs in binaries 
is the thermal--viscous instability driven by hydrogen 
ionization (Meyer \& Meyer--Hofmeister 1982, Smak 1982). It is now
commonly accepted that this instability 
drives outbursts in cataclysmic variables and soft X--ray transients. 
Lin \& Shields (1986) showed by a local stability analysis that
this instability can also operate in accretion
discs thought to be present around supermassive black holes in AGN. 
They concluded that these discs were unstable
at radii ($ \approx 10^{15} - 10^{16}$ cm), where the surface 
temperature is several thousand degrees. The expected characteristic 
time scale for this instability is $10^4 - 10^7$ years.

Because of its generic nature, the ionization instability 
plays a dominant role in characterizing the observed behaviour of the
host systems. In the binary context, attempts to understand the  
precise conditions (mass of accreting object, accretion rate) under
which it occurs have been at least partially successful
(e.g. Smak, 1982, van Paradijs 1996; King, Kolb, \& Burderi, 1996;
King, Kolb \&
Szuszkiewicz, 1997, and references therein). These studies show that
self--irradiation of the disc by the central X--ray source has a
determining effect on the disc stability if the accreting object is
compact, as in soft X--ray transients (see below).
Delineating the stable and unstable disc regions is equally important
for AGN. If the instability is present in AGN discs the
suppression of central accretion in the 
quiescent state means that we can identify
only the outburst states of unstable systems as AGN. Two important
consequences follow: (1) quiescent AGN must appear as quite normal
galaxies, and (2) the average mass fuelling rate in many if not all
AGN is much lower
than implied by their current luminosities. This in turn limits the
masses that their central black holes are expected to reach.
	

The idea of intermittent activity in AGN was already suggested 
by Shields \& Wheeler (1978). They noticed that the fuelling problem
could  be solved if active nuclei
store mass during quiescence and this mass then 
feeds  the hole for a shorter period of intense 
activity. The thermal--viscous hydrogen  ionization instability found to
operate in AGN accretion discs (Lin \& Shields, 1986) 
is capable of triggering such  behaviour.  
Clarke \& Shields (1989), Mineshige \& Shields (1990), 
Cannizzo \& Reiff, 1992, Cannizzo (1992)
studied the full range of black hole masses and accretion rates in order to
determine the observational consequence of the instability for the AGN
population. Siemiginowska \& Elvis (1997)
attempted to reproduce the observed luminosity function, assuming
that this mechanism operates in all AGN.

Our aim here is to decide if the ionization instability
still operates in AGN when irradiation effects are included.
As we have seen irradiation is central to the discussion of disc
stability in soft X--ray transients. Further,
irradiation is
often thought to dominate the disc emission (e.g. Collin--Souffrin, 1994). 
For both reasons it is vital to include it in any attempt to
decide the disc stability. 
The actual form of the instability when irradiation is included 
is outside the scope of our paper. Siemiginowska, Czerny \& Kostyunin (1996)
have performed studies for particular 
black hole masses and accretion rates, with assumed forms of irradiation.


There is a simple criterion for the instability to appear: the disc
must contain regions with effective temperature $T_{\rm eff}$ 
close to the value
$T_{\rm H}$ at which hydrogen is ionized. In practice $T_{\rm H}$ 
depends on the density and may be quite different in different
environments; 
 we shall consider a
range of values in this paper. However the criterion is not easy
to use in this form, as one does not in general know the radial
distribution of the accretion rate, and thus the run of $T_{\rm eff}$,
in a time--varying disc. Accordingly one usually uses the criterion in an
indirect form: a disc with a given constant accretion rate $\dot M$ is
self--consistently steady only if $T_{\rm eff} > T_{\rm H}$ throughout
it. If the criterion fails we may expect outbursts, although the
precise nature of these will depend for example on the detailed behaviour of
the disc viscosity. 

This version of the stability criterion is easy to apply. Since
$T_{\rm eff}$ always decreases with disc radius $R$ in a steady disc,
the condition is most stringent at the outer disc radius $R_{\rm
out}$, so we need apply it only there. If the disc's only source of
energy is local viscous dissipation we have 
\begin{equation}
%
%
\left[ T_{\rm eff}(R) \right] ^4 = {3GM\dot{M}\over 8\pi R^3\sigma}f ,
\label{eqa}
\end{equation}
(e.g. Frank, King \& Raine (1992); all symbols are explained after 
equations (2--9)),
and the criterion is simply 
$T_{\rm eff}(R_{\rm out}) > T_{\rm H}$. In a binary system we can
estimate $R_{\rm out}$ with reasonable accuracy as 70\% of the Roche
lobe radius of the accreting star, and the problem is now well
determined. Using this approach, Smak (1982) successfully divided
outbursting cataclysmic variables (dwarf novae) from the persistent
systems (novalikes). The extension to low--mass
X--ray binaries is complicated by the fact that the dominant heat
source for the disc is not local viscous dissipation (equation  (\ref{eqa})),
but irradiation by the central X--rays. The instability is similarly
suppressed if the disc surface temperature given by irradiation
exceeds $T_{\rm H}$ (Tuchman, Mineshige \&
Wheeler, 1990). Provided that
due account is taken of this, one can again successfully divide the
outbursting systems (soft X--ray transients) from the persistent
systems (van Paradijs, 1996; King, Kolb \& Szuszkiewicz, 1997). The
key feature, as in the unirradiated case, is that the edge temperature
of the disc can be simply expressed in terms of $\dot M$ and 
$R_{\rm out}$, without any need to solve for the full internal disc structure.
In both the CV and LMXB cases there are important consequences for the
study of the binary evolution (e.g. King, Kolb \& Burderi, 1996),
which gives a connection between $\dot M, M$ and $R_{\rm out}$. 

The extension of this approach to AGN is more complicated; here the
outer edge of the disc is no longer determined by the simple Roche
lobe condition which holds in binaries, but by the requirement that
the disc becomes locally self--gravitating (see equation  (\ref{13}) below).
This condition requires a knowledge of the disc density at the
outer edge, so we are now required to solve the full global structure of
the steady disc to find $R_{\rm out}$. 
 Thus we examine all possible, stationary,
optically thick, geometrically thin disc models relevant for AGN. If
these correspond to stable states, AGN discs may be globally steady,
and require fuelling at the currently inferred central accretion
rates. If not, they will be the outburst states, and the required
fuelling rates will be lower than the current central accretion rate.

\section{GLOBAL DISC STRUCTURE}

As explained above,
to apply the stability criterion we need the self--gravity radius, and thus
the global structure of steady discs. We assume these to be optically thick 
and geometrically thin. If we exclude the region
close to the central object and consider total disc luminosities
$L \lta 0.2L_{\rm E}$ ($L_{\rm E}$ being the Eddington luminosity)
these approximations are justified, and we may
parametrize all possible disc structures by $\dot M$,
$M$ and the viscosity parameter $\alpha$ (see below). 
Our approach exploits only hot radiative regime of the disc,
for which vertically average structure is a good approximation.   
The algebraic system describing a Shakura--Sunyaev disc is (see
{\it e.g.} Frank, King, \& Raine, 1992, p. 77):
\begin{equation}
\rho = \Sigma H^{-1} 
\label{eq1}
\end{equation}

\begin{equation}
H = c_{\rm s} R^{\frac{3}{2}} G^{-\frac{1}{2}} M^{-\frac{1}{2}} \\
\end{equation}

\begin{equation}
c_{\rm s} = P^{\frac{1}{2}} \rho^{-\frac{1}{2}} 
\end{equation}

\begin{equation}
P = \frac{k}{\mu m_{\rm p}} \rho T + \frac{4 \sigma}{3 c} T^4 
\end{equation}

\begin{equation}
{ 4\sigma T^4 \over 3 \tau }  = \frac{3GM\dot{M}}{8\pi R^3}f +
Q_{\rm irr}  
\end{equation}

\begin{equation}
\tau = \Sigma \kappa 
\end{equation}

\begin{equation}
\Sigma = \frac{1}{3 \pi} \dot{M} \nu^{-1}f
\end{equation}

\begin{eqnarray}
\label{eq8} 
\nu = \frac{2}{3} \alpha H c_{\rm s} \ \ \ {\rm for \ \ \alpha}{\rm-discs} \\ 
\nonumber
\\ \nonumber 
\nu = \frac{2}{3} \alpha H c_{\rm s}\left({P_{\rm g} \over P}\right) \ \ \  
{\rm for \ \ \beta}{\rm-discs}.
\end{eqnarray}
Here $G$ is the gravitational constant, $R$ is the radial
distance from the central object (assumed not to be extremely close to
the inner radius of the disc), $\Sigma$ is the surface density of
the disc, $\kappa = \kappa_{\rm abs} + \kappa_{\rm es}$ is the opacity 
($\kappa_{\rm abs}$ being the part due to (mainly) free--free and
free--bound processes, and $\kappa_{\rm es}$ being the part due to
electron scattering), $H$ is the disc half--thickness, $c_{\rm s}$ is the sound
speed, 
%
%
$\rho$ is the density of the disc, $T$ is its central temperature related
to the effective temperature by $T^4 = (3\tau/4) T_{\rm eff}^4$,
$Q_{\rm irr}$ is the flux absorbed from the external radiation (see equation
(14)), $M$ is the mass of the black hole, $\dot{M}$ is the accretion rate,  
%
%
$P = P_{\rm rad} + P_{\rm g}$ is the total pressure 
($P_{\rm rad}$ and $P_{\rm g}$ being the radiation pressure and the
gas pressure respectively),
$\tau$ is the optical depth, $\nu$ is the
kinematical  viscosity, $k$ is the Boltzmann constant,
$m_{\rm p}$ is the mass of the proton, $\mu$ is the mean particle mass
of the accreting material, $\sigma$ is the Stefan--Boltzmann constant,
$c$ is the speed of light, $f= [1-(6GM/c^2R)^{1/2}]\approx 1$, as we
are interested in the outer regions of the discs 
and $\alpha$ is the Shakura--Sunyaev
parametrisation of the viscosity.
Equation (9) gives  two different viscosity prescriptions. The first 
corresponds
to a  situation where a given viscous stress tensor component 
 is proportional to
the total pressure (discs with this viscosity prescriptions are called 
$\alpha$--discs) 
and the second when it is proportional to the gas pressure only (discs with
this viscosity prescription are called $\beta$--discs). 

\subsection {DISC WITH NO IRRADIATION} 
In general we can divide an optically thick  disc into four different regimes,
namely 
$$ P_{\rm g} \ll P_{\rm rad}, \ \ \kappa_{\rm es} \gg \kappa_{\rm abs } 
\ \ \rm inner \ \ 
 region \ ({\it a})$$
$$ P_{\rm g} \gg P_{\rm rad}, \ \ \kappa_{\rm es} \gg \kappa_{\rm abs } 
\ \ \rm middle \ \ 
 region \ ({\it b}) $$
$$ P_{\rm g} \ll P_{\rm rad}, \ \ \kappa_{\rm es} \ll \kappa_{\rm abs } 
\ \ \rm middle^*
\ \ region \ ({\it b^*})
$$
$$ P_{\rm g} \gg  P_{\rm rad}, \ \ \kappa_{\rm es} \ll \kappa_{\rm abs} 
\ \ \rm outer \ \ 
 region \ ({\it c})$$

The names of the regions correspond to those used by Shakura \& 
Sunyaev (1973); region {\it b$^*$}  was not included in their paper. 
We have found
that this region exists   in  discs with $\alpha \gta 3\times 10^{-3} $. 
Which regions are actually present in a given disc 
 depends on $M$, $\dot M$,
and $\alpha $. 

Shakura \& Sunyaev used only the Rosseland mean for 
free--free absorption
\begin{equation}
\kappa_{\rm abs } = 6.2\times 10^{22} \rho T^{-3.5}\ {\rm cm^2g^{-1}}.
\label{kap}
\end{equation}
However for most relevant densities and temperatures bound--free opacities
contribute significantly to the total, and this approximation is quite
inadequate. 
Some consequences of using (\ref{kap}) have been pointed out by
Cannizzo \&  Reiff (1992) and Hur\'e et al. (1994b).
In fact many {\it local} considerations  are unaffected
by the change in opacities, as most quantities depend only
very weakly on them (typically as $\sim \kappa_{\rm abs }^{0.1}$). 
However the {\it global} structure of the disc is very severely
affected by the use of the wrong opacities, as these determine where
the various regions {\it a -- c} match to each other. 
This is illustrated in Figure 1 where the borders $R_{ab}$, $R_{bc}$
between regions $a$ and $b$, 
and between $b$ and $c$, are given as functions 
of $\dot M$ for two different opacity approximations. The mass of
the central black hole is taken equal to $10^8 M_{\odot}$ and 
parameter $\alpha =
0.001$. The solid line
for $R_{bc}$ is obtained 
using a power law fit to the 
recently compiled solar--abundance opacities 
(Mazzitelli, 1989 and references therein) in the relevant
temperature--density regime. 
\begin{equation}
\kappa_{\rm abs } = 9\times 10^{24} \rho T^{-3.5}\ {\rm cm^2g^{-1}}.
\label{11} 
\end{equation}
The dotted line gives to the $R_{bc}$ values if  the opacities 
are given by equation (10). The filled squares are taken from
the numerical calculation (Szuszkiewicz, Malkan \& Abramowicz 1996)
where Cox \& Stewart (1970) opacity
tables were used. The values  of $R_{ab}$ are  the
same in both cases, as in regions $a$ and $b$ electron scattering is a 
dominant source of opacity. We list all relevant expressions in Appendix A. 

The disc is terminated at the radius where self--gravity
becomes important. This is evaluated using the stability criterion for
a differentially rotating disc (Toomre 1964)
\begin{equation}
Q_{\rm T} = c_{\rm s} \Omega/\pi G \Sigma \gg  1
\end{equation}
where $\Omega$ is the angular velocity.
The condition $Q_{\rm T}=1$ defines a self--gravity radius
\begin{equation}
R_{\rm sg} = (M/\pi \rho)^{1/3}.
\label{13} 
\end{equation}
Knowing the gas density from the given accretion disc
model we can find the  outer edge of the disc.  
The disc can terminate in any of the regions mentioned above and the 
appropriate
density structure must be used. 
After simple algebra one can find appropriate formula for the self--gravity
radius for every region in the disc. We give them for all regions and
for both $\alpha$-- and $\beta$--discs in Appendix C.  
We compared our results with the self--gravity radii calculated by 
Hur\'e et al. (1994a) and found reasonable agreement. 
Many formulae for the self--gravity radius
used in the literature simply assume  that this radius
is always located in region {\it a}, although there is no particular reason
to believe this.
Region {\it a} can also be terminated by the change from radiation
to gas pressure before the density reaches the  self--gravitating value.
In order to determine in which region, $a$, $b$, $b^*$ or  $c$ a disc
actually ends it is sufficient to compare  
the dimension of each region with the appropriate self--gravity
radius for given $M$, $\dot M$ and 
$\alpha$. The results are shown in Figure 2. 

\subsection{IRRADIATED DISCS}

If irradiation dominates the heating, the disc
can extend to much larger radii before self--gravity starts to be important. 
A similar case was considered by Ko \& Kallman (1991), but their results
are affected by the use of the opacity form (\ref{kap}). 
The structure of the irradiated  disc can be found from
 equations (\ref{eq1} -- \ref{eq8}).   
Viscous energy generation is now negligible in comparison with
the flux absorbed from the external radiation, $Q_{\rm irr}$.  
For photons emitted radially from a
central source, 
$Q_{\rm irr}$ has the form
\begin{equation}
Q_{\rm irr}  = \frac{\eta \dot{M}c^2(1- \chi )}{4\pi R^2} R \frac{d(H/R)}{dR} 
\label{qirr}
\end{equation}
where $\eta \;\;(\sim 0.1)$ is the efficiency of the accretion process,
$\chi $ (reasonable values are between 0.1 to 0.9) is the albedo of 
the disc and $ R \frac{d(H/R)}{dR}$ takes
into account the projected surface area of the disc normal to the
radiation flux. If the central emission is not radial but from the
inner disc $Q_{\rm irr}$ is multiplied by a factor $H/R$ because this
source is foreshortened; cf King, Kolb \& Szuszkiewicz, (1997). We
shall not consider this case here, because we shall find that 
even with the favourable
assumption (\ref{qirr}) irradiation is never crucial in deciding the
{\it stability} 
of an AGN disc. 
%
%
We strongly  suspect that this is true  also for 
other possible ways 
in which a disc can be illuminated  (for example by 
an extended corona or a jet). However 
to show this requires additional
assumptions about these poorly--understood configurations
and we leave this for further investigation.  
For low luminosities and low central masses the disc structure is
altered by the presence of irradiation.  
This is shown in
Figure 2, where the part of the $\dot M-M$ plane for which the
disc is terminated in an irradiated region is called C$^+$. The thickness
of the disc in the irradiated part does not vary with radius as
$H \propto R^{9/8}$ (the non--irradiated case) but as $H \propto R^{45/38}$.
Irradiation effects are relevant in zone {\it c} if

\begin{equation}
{L \over L_{\rm Edd}}={\dot M  \over \dot M_{\rm c}} \lta
7\times 10^{-5}   \alpha ^{3\over 2} 
M _8 ^{-\frac {7}{2}} (1-\chi)_{0.9}^{\frac {5}{2}}
\label{15}
\end{equation} 

in zone  $b$
if 

\begin{equation}
{\dot M  \over \dot M_{\rm c}} \lta
1\times 10^{-16}   \alpha ^{4} 
%
%
M_8 ^{-10} 
(1-\chi)_{0.9}^{9} 
\label{16}
\end{equation} 

in zone $b^*$ (for $\alpha $--discs) if 

\begin{equation}
{\dot M  \over \dot M_{\rm c}} \gta
3\times 10^{2}   \alpha ^{-1} 
M_8  ^{3 \over 2}
(1-\chi)_{0.9}^{-\frac {5}{2}}
\label{17}
\end{equation} 
 
and in zone $b^*$ (for $\beta $--discs) if

\begin{equation}
{\dot M  \over \dot M_{\rm c}} \gta
1\times 10^{7}   \alpha ^{-3}
M_8  ^{11 \over 2}
(1-\chi)_{0.9}^{-\frac {13}{2}}
\label{18}
\end{equation}

Here  $L_{\rm Edd}=1.5\times 10^{46}M_8$ erg s$^{-1}$  
is the Eddington luminosity, 
$\dot M_{\rm c}=2.6\times 10^{26}M_8$ g s$^{-1}$,
$M_8= M/10^8M_{\odot}$ and $(1-\chi)_{0.9}={\frac {1-\chi}{0.1}}$. 
 The thin disc approximation we use here requires
$\dot M /\dot M_{\rm c} \lta 0.2$.
The vertical thickness of the disc in
region {\it a} is constant and it therefore cannot be irradiated by photons
emitted radially from a central source, as it
is all shadowed (cf equation  (\ref{qirr})). The situation will be, of course,
quite different in the case of irradiation by the corona or emission
scattered above the disc.

\section{DISC STABILITY}

Armed with the values of the disc self--gravity radius from the
previous sections, we can now check the simple criterion
$T_{\rm eff}(R_{\rm sg}) > T_{\rm H}$ for stable disc accretion. We
consider values $T_{\rm H} = 6500, 2000$ K, corresponding to the
extremes of what is normally claimed for AGN discs
(Lin \& Shields, 1986; Clarke, 1988; Clarke \& Shields, 1989; 
 Mineshige \& Shields, 1990; Cannizzo,  1992).

Accordingly we identify the stable regimes in  
regions A, B, B$^*$ C, and C$^+$ for both $\alpha$ and $\beta$ discs 
(Figure 2). Here the $\dot M -M$ plane  
is divided into regions A, B, or B$^*$, C, and C$^+$
%
%
consisting of all models in which the disc is terminated
by self--gravity in regions {\it a, b, $b^*$, c }  and {\it c$^+$} respectively. 
The Eddington limit and the limit of validity for the thin disc approximation,
namely $L \lta 0.2L_{\rm E}$,   
%
%
are also shown.

Unlike in the case of low--mass X--ray binaries,
 even in the extreme case of irradiation 
with very low albedo (90\% of the incident radiation absorbed by
the disc), irradiation never stabilizes the disc.
To see this, we note
that the criterion for stability in the irradiated disc in zone $c$ is
\begin{equation}
7\times 10^{-5} \alpha ^{3\over 2} M_8
^{-{7\over 2}} 
(1-\chi)_{0.9}^{\frac {5}{2}}
 \gta 
{\dot M  \over \dot M_{\rm c}} 
\gta 2.6  \alpha ^{17\over 26} 
M_8
^{-{17\over 26}}
(1-\chi)_{0.9}^{-\frac {9}{26}}
\left[T_{\rm H}\right]^{24\over 13}
_{2000} 
\end{equation}
where $\left[T_{\rm H}\right]_{2000}$ is $T_{\rm H}/2000$.
From this criterion and from the requirement that  $\dot M/\dot M_{\rm c} 
 \lta 0.2$  it 
follows easily that irradiation is unable to stabilize the disc
in zone $c$. Equivalent criteria for zone $b$, $b^*$
lead to the same conclusion. 


\section{DISCUSSION}

Our aim in this study was to investigate whether the thermal--viscous
 ionization 
instability operates in AGN  in the presence of irradiation.
We have studied
stationary, optically thick, geometrically thin discs,
in the range of accretion rates and central black hole 
masses for which these models are self--consistent. It is worth mentioning
here that advection dominated optically thin discs can in principle
coexist in some particular regions of the parameter space, but which
type of the solution will be actually chosen in nature is still an open
question. 
We used a very simple analytic criterion to determine the stability of each
model; if the disc is hot enough for  hydrogen to be completely
ionized everywhere all way out till its self--gravity radius 
the ionization  instability cannot operate. We
identify such hot regions in the relevant parts  of $\dot{M}$ -- $M$  
%
%
plane and
show them as  grey (for $\alpha$--discs) and hatched (for $\beta$--discs)  
areas in Figure 2. Unlike other authors 
(Clarke \& Shields, 1989; Mineshige \& Shields, 1990,
Cannizzo,  1992)
we consider only the
upper stable branch of the whole cycle, where our method is appropriate.
A major advantage of our approach is that  we do 
not need a complicated discussion of the  limit cycle. 
This method  proved successful in  similar studies 
of accretion discs in X--ray binaries. 
We gave careful consideration to the opacities used in  our calculations. 
There are only small differences between results
using opacities from
Mazzitelli (1989) and Cox \& Stewart (1970).
However differences appear when using  simple
fitting  formulae such as  (10) instead of (11) (see Figure 1): 
it is important to check carefully that a particular fit found in
the literature is appropriate for the range of temperatures and densities
used in a given  problem.

Another result of our study is
that for $\alpha \gta 0.003$ the region 
between region $a$ and $c$ differs from the standard Shakura--Sunyaev
region $b$. We denote it $b^*$. It is radiation pressure dominated, but
the main source of opacity is true absorption. We have confirmed
the existence of this 
region in numerical calculations of global
disc structure performed using the Cox \& Stewart
(1970) opacity tables. It is interesting that $b^*$ is stable
against disc instabilities  triggered by  radiation pressure (Pringle, 1976): 
while irradiated  it might significantly change its
properties. 

In Figure 3 we compare our results with those based on detailed studies
of the outburst cycle over the parameter space considered 
by various authors.
The dotted lines are 
from Mineshige \& Shields (1990), dotted--dashed from Clarke \& Shields (1989),
long--dashed from Cannizzo (1992) and the bold lines from  this paper. 
The short--dashed line gives the Eddington limit. 
Our results
for non--irradiated disc are in good agreement with those obtained
previously. Our main result, quite contrary to the case of close
binaries, is
that irradiation does  not change the  borders between unstable and stable
(partially or completely ionized) regions. In other words, irradiation
by a central point source is unable to stabilize the whole disc out to its
self--gravity radius. 
An important reason for this is that one of the effects  of
such irradiation is to move the self--gravity radius even farther out from
the central black hole. The irradiated disc structure for low--luminosity,
low--mass objects differs from 
that of the  equivalent discs without irradiation
(regions C$^+$ in Figure 2). Thus
the actual appearance of the ionization instability might 
 well  be
affected. This can be studied
%
%
only by detailed calculations of thermal
limit cycles in
%
%
the presence of irradiation.

We see from Figure 2 that in general AGN discs will be subject to the
ionization instability, even if they are irradiated by a central point
source. 
 For typical AGN luminosities,
corresponding to central accretion rates $\lta
10^{-2}M_{\odot}$~yr$^{-1}$, we see from Figure 2 that it is inconsistent to
assume that the disc is stable.
Since central accretion (and thus e.g. X--ray emission)
is suppressed in the
quiescent state, all observed 
%
%
AGN must presumably be identified as such only 
in their outburst states (which last $\gta 10^{3}$ yr). Thus AGN currently
observed to have central accretion rates below the stability limits 
$\sim 10^{-1} - 10^{-2}M_{\odot}$ yr$^{-1}$ shown in Figure 2
must actually have considerably
lower fuelling rates. Even rather brighter observed AGN need not be
steady systems, but may simply represent the outburst states of
unstable disc with fuelling rates below the stability limits.

As pointed out
in the Introduction, if most AGN discs are unstable, then in quiescence these
systems must be
indistinguishable from normal galaxies. Moreover the mass fuelling
rates needed to power AGN must be much lower than implied by their current
luminosities. If the duty cycle for the outburst
can be made short enough ($\lta 10^{-2}$), no fuelling rates greater than about
$10^{-2}\ M_{\odot}$~yr$^{-1}$ would be needed in AGN. This would also
remove the problem that the remnant black holes are predicted to have
excessively high masses if accretion is continuous (Cavaliere \&
Padovani 1988).

Alternatively, since most
quasars have observed central accretion rates above the
stability limits in Figure 2, some or all of them
could have steady discs. This group would then form
a separate class with much higher fuelling rates 
$\dot M \sim 0.1 - 1\ M_{\odot}$~yr$^{-1}$. 
It is not easy to decide between
these possibilities by looking at detailed properties of the individual 
systems, as outbursting discs rapidly take on a quasi--steady surface density
profile (cf Cannizzo, 1993: this property is well known in the context of
cataclysmic variables, where the persistent systems -- novalike
variables -- look like dwarf novae in permanent outburst). A complicating
feature is that many of the objects with high steady fuelling rates
would be subject to the radiation--pressure (Lightman--Eardley) instability.

We conclude that many (if not all) AGN represent the outburst
state of a thermal--viscous disc instability. We should then consider
candidates for the quiescent state. It is tempting to suggest that
this may comprise most or all ``normal'' spirals. Galaxies such as our
own could therefore harbour moderately massive ($10^6 -10^8M_{\odot}$)
black holes in their nuclei.

{\bf Acknowledgements} We thank Ulrich Kolb for valuable discussions.  
 This work is supported by a PPARC Rolling Grant for 
theoretical astrophysics to the University of Leicester. ARK gratefully
acknowledges the support of a PPARC Senior Fellowship.

\clearpage
  
\section*{FIGURE CAPTIONS}
\bigskip

\figcaption{The radii $R_{ ab}$, $R_{ bc}$ (in terms of the Schwarzschild
radius, $r_g=2GM/c^2$)
dividing regions $a$ and $b$,  and regions 
$b$
and $c$, as functions of  accretion rate (in terms of the critical accretion
rate, $\dot M_c$). 
$R_{bc}$ shown
by the solid line was obtained using opacities given by equation (11) and
that shown by the dotted line was calculated with opacities given by 
equation (10). 
$R_{ab}$ are  the same in both cases. Here $M=10^8M_{\odot}$
and $\alpha =0.001$. 
The filled squares are taken from
the numerical calculation (Szuszkiewicz, Malkan \& Abramowicz 1996)
where Cox \& Stewart (1970) opacity
tables were used.} 
\figcaption{
For a given set of parameters ($\alpha$, $\dot M$ and $M$) the global disc
structure is uniquely determined. Region A contains all models in which
self--gravity truncates the disc in a region where
the pressure  is dominated by radiation and the opacity by
electron scattering (Shakura--Sunyaev region $a$). 
Region B contains the models in which the self--gravity radius terminates the
disc in the zone where the pressure is dominated by gas pressure and 
opacity by electron scattering (Shakura--Sunyaev region $b$). 
Region B$^*$ is the regime 
in which the self--gravity radius occurs in the zone where
radiation pressure dominates, but the main source of opacity is 
absorption rather than scattering. 
Region C contains models with the outer radii determined
by self--gravity in the zone where gas pressure dominates
and opacity is given by absorption (Shakura--Sunyaev region $c$). 
 In all regions A, B, B$^*$ and C
the main source of heating is due to the viscous energy dissipation.
In region C$^+$ the main heating process is irradiation by
the central source instead. The size of this region depends on the 
albedo of the disc, $\chi  $. It is shown here 
 for two particular values of $\chi $:
0.1 and 0.9. 
%
%
%
%
In every region the stability properties have been determined,  both for 
$\alpha$ and $\beta$--discs,  for
two  values of hydrogen ionization temperature $T_H$ (left panel:
$T_H = 6500$ K, right panel: $T_H =2000$ K) and for different values of
$\alpha$ parameter (left panel: 
$\alpha =10^{-4}$, $\alpha =10^{-3}$, $\alpha =10^{-2}$,  
right  panel: 
$\alpha =10^{-3}$, $\alpha =10^{-2}$, $\alpha =10^{-1}$).
The shaded zones show the
location of the $\alpha$--discs models in which the ionization
instability does not operate. The hatched zones on the left panel show the
location of the $\beta $--disc models in which the ionization
instability is suppressed. For $T_H=2000$ K (right panel) 
 the stability properties
for $\alpha$ and $\beta$--discs are the same, so the shading shows the
stable region for both cases.}

\figcaption{$\dot{M}$ -- $M$
%
%
plane for AGN discs, divided into
three distinct regions according to the degree of hydrogen ionization:  
 neutral, partially ionized
and completely  ionized. We take $\alpha = 0.1 $. The box with
 thick solid lines defines the domain discussed in this paper, and
horizontal line inside it shows the accretion rate above which the
discs are completely  ionized. Previous studies of the
 $\dot{M}$ -- $M$ 
%
%
plane
are shown by different lines: Clarke \& Shields (1989) -- dot--dashed,
Mineshige \& Shields (1990) -- dotted, Cannizzo (1992) -- long--dashed.
The short--dashed line represents the Eddington limit. 
See text for full details.}
\bigskip
\newpage 

\section*{APPENDIX A}
\begin{center}
{\bf  Size of the regions  in non--irradiated  $\alpha$ and $\beta$--discs }
\end{center}

Here and in Appendices B and C  we give  expressions for all characteristic
radii used in our paper. They are given in a form  of two equalities. In the
first equality, for generality, we retain a dependence on
the electron scattering opacity coefficient
$\kappa_{\rm es} =0.2(1+X)$ [cm$^{2}$/g], where $X$ is 
 the hydrogen content by mass (in this paper we use  X=0.7), and 
on  $\kappa_{\rm o}$ 
which  is a constant 
coefficient used in our  fitting procedure: 
$$\kappa_{\rm abs} =\kappa_{\rm o}\rho T^{-3.5}  [{\rm cm}^2/{\rm g}]$$
The best fit to the solar--abundance opacities (Mazzitelli, 1989) was  
 obtained with   $\kappa_{\rm o}= 9\times 10^{24}$. 
Moreover, still in the first equality , $\dot M$ should be  in [g/s] 
and $M$  in [g] in order to get the radii in [cm]. In the second equality,
in order to make the use of our formulae easier,  we introduced the following
quantities: $$\dot M_{-1} = { \dot M \over 0.1\dot M_{c}}  \ \ {\rm and}
\ \ \ M_8 ={M \over 10^8M_{\odot}}$$
where $\dot M_c =2.6\times 10^{26} M_8$ [g/s]. All the coefficients  
 in front  of these
quantities are given in [cm]. 
 
 Note that for $\alpha $ less than 0.003 only region $b$ is present in a disc, 
 and for $\alpha $ greater than 0.003 only region $b^*$ is present. 

\bigskip
\begin{equation}
\begin{array}{l}
R_{bc}= 2.8224 \times 10^{2} 
\kappa_{\rm o}^{-\frac{2}{3}}
\kappa_{\rm es}^{\frac{4}{3}}
\dot{M}^{\frac{2}{3}}\
M^{\frac{1}{3}} 
\\

\\
\ \ \ \ \ =
7.9 \times 10^{15} \dot M_{-1}^{\frac{2}{3}} M_8 
\\

\\

\\
R_{b^*c}= 1.5089 \times 10^{-51} 
\kappa_{\rm o}^{\frac{6}{5}}
\alpha ^{\frac{4}{15}}
\dot{M}^{\frac{14}{15}}\
M^{\frac{1}{3}} 
\\

\\
\ \ \ \ \ \ = 
4.1 \times 10^{16}\alpha ^{\frac{4}{15}} \dot M_{-1}^{\frac{14}{15}} 
M_8^{\frac{19}{15}}
\\

\\

\\
R_{ab}= 2.6602 \times 10^{-17} 
\kappa_{\rm es}^{\frac{18}{21}}
\alpha^{\frac{2}{21}}
\dot{M}^{\frac{16}{21}}\
M^{\frac{1}{3}} 
\\

\\
\ \ \ \ \ = 
1.4 \times 10^{16}\alpha ^{\frac{2}{21}} \dot M_{-1}^{\frac{16}{21}} 
M_8^{\frac{23}{21}}
\\

\\

\\
\hspace{-1.5cm}
{\rm for \ \alpha-discs }
\\

\\
R_{ab^*}= 3.7322\times 10^{-7} 
\kappa_{\rm o}^{-\frac{16}{45}}
\kappa_{\rm es}^{\frac{10}{9}}
\alpha^{\frac{2}{45}}
\dot{M}^{\frac{32}{45}}\
M^{\frac{1}{3}} 
\\

\\
\ \ \ \ \ \ \ = 
1.0 \times 10^{16}\alpha ^{\frac{2}{45}} \dot M_{-1}^{\frac{32}{45}} 
M_8^{\frac{47}{45}}
\\

\\
\hspace{-1.5cm}
{\rm for \ \beta-discs }
\\
 
\\
R_{ab^*}= 4.1353\times 10^{-6} 
\kappa_{\rm es}^{\frac{58}{51}}
\kappa_{\rm o}^{-\frac{20}{51}}
\alpha ^{\frac{2}{51}}
\dot{M}^{\frac{36}{51}}\
M^{\frac{1}{3}} 
\\

\\
\ \ \ \ \ \ \ = 
1.0 \times 10^{16}\alpha ^{\frac{2}{51}} \dot M_{-1}^{\frac{36}{51}} 
M_8^{\frac{53}{51}}
\end{array}
\end{equation}

\newpage 

\section*{APPENDIX B}
\begin{center}
{\bf  Size of the regions {\it c} and {\it c}$^+$, {\it b} and {\it b}$^+$
and $b^*$ and $b^{*+}$ of irradiated discs }
\end{center}
 
\bigskip
\begin{equation}
\begin{array}{l}
R_{cc^+}= 2.5706 \times 10^{-28}
\kappa_{\rm o}^{-\frac{2}{45}}
\alpha^{\frac{4}{45}}
\dot{M}^{-\frac{2}{15}}\
M^{\frac{11}{9}} 
\\

\\
\ \ \ \ \ \ \ =
2.4 \times 10^{18}\alpha ^{\frac{4}{45}} \dot M_{-1}^{-\frac{2}{15}} 
M_8^{\frac{49}{45}}
\\

\\

\\
R_{bb^+}= 2.0215 \times 10^{-30}
\kappa_{\rm es}^{-\frac{2}{21}}
\alpha^{\frac{2}{21}}
\dot{M}^{-\frac{4}{21}}\
M^{\frac{9}{7}} 
\\

\\
\ \ \ \ \ \ \ =
4.1 \times 10^{18}\alpha ^{\frac{2}{21}} \dot M_{-1}^{-\frac{4}{21}} 
M_8^{\frac{23}{21}}
\\
 
\\
  
\\
\hspace{-1.5cm}
{\rm for \ \alpha-discs }
\\
 
\\
R_{ b^*b^{*+}}= 8.8331 \times 10^{-23} 
\kappa_{\rm o}^{-\frac{16}{45}}
\alpha^{\frac{2}{45}}
\dot{M}^{-\frac{2}{5}}\
M^{\frac{13}{9}} 
\\

\\
\ \ \ \ \ \ \ \ \  =
3.6 \times 10^{18}\alpha ^{\frac{2}{45}} \dot M_{-1}^{-\frac{2}{5}} 
M_8^{\frac{47}{45}}
\\
 
\\
\hspace{-1.5cm}
{\rm for \ \beta-discs }
\\
 
\\
R_{ b^*b^{*+}}= 3.9392 \times 10^{-22}
\kappa_{\rm o}^{-\frac{20}{51}}
\alpha^{\frac{2}{51}}
\dot{M}^{-\frac{22}{51}}\
M^{\frac{25}{17}} 
\\

\\
\ \ \ \ \ \ \ \ \ =
3.8 \times 10^{18}\alpha ^{\frac{2}{51}} \dot M_{-1}^{-\frac{22}{51}} 
M_8^{\frac{53}{51}}
 
\end{array}
\end{equation}

\newpage 
\section*{APPENDIX C}
\begin{center}
{\bf Self--gravity radius for each region in non--irradiated and irradiated 
 discs  }
\end{center}
\begin{center}
{\bf Region  {\it c} (for both $\alpha$ and $\beta$--discs) }
\end{center}
\bigskip
\begin{equation}
\begin{array}{l}
R_{\rm sg}= 3.9687\times 10^{12} 
\kappa_{\rm o}^{\frac{2}{15}}
\alpha^{\frac{28}{45}}
\dot{M}^{-\frac{22}{45}} \
M^{\frac{1}{3}}
\\

\\
\ \ \ \ \  \ =
1.8 \times 10^{17}\alpha ^{\frac{28}{45}} \dot M_{-1}^{-\frac{22}{45}} 
M_8^{-\frac{7}{45}}
\end{array}
\end{equation}
\bigskip
\begin{center}
{\bf Region  {\it b} (for both $\alpha$ and $\beta$--discs) }
\end{center}
\begin{equation}
\begin{array}{l}
R_{\rm sg}= 8.0781\times 10^{10} 
\kappa_{\rm es}^{\frac{2}{9}}
\alpha^{\frac{14}{27}}
\dot{M}^{-\frac{8}{27}} \
M^{\frac{1}{3}} 
\\

\\
\ \ \ \ \  \ =
1.1 \times 10^{17}\alpha ^{\frac{14}{27}} \dot M_{-1}^{-\frac{8}{27}} 
M_8^{\frac{1}{27}}
\end{array}
\end{equation}
\bigskip
\begin{center}
{\bf  Region  {\it b}$^*$ for $\alpha $--discs}
\end{center}
\begin{equation}
\begin{array}{l}
R_{\rm sg}  =  6.5485 \times 10^{-12} 
\kappa_{\rm o}^{\frac{8}{15}} \ \
\alpha^{\frac{22}{45}} \
\dot{M}^{\frac{2}{45}} \
M^{\frac{1}{3}} 
\\

\\
\ \ \ \ \  \ =
1.0 \times 10^{17}\alpha ^{\frac{22}{45}} \dot M_{-1}^{\frac{2}{45}} 
M_8^{\frac{17}{45}}
\end{array}
\end{equation}
\bigskip
\begin{center}
{\bf  Region  {\it b}$^*$ for $\beta$--discs}
\end{center}
\begin{equation}
\begin{array}{l}
R_{\rm sg}  = 1.7057 \times 10^{2} 
\kappa_{\rm o}^{\frac{4}{13}} \ \
\alpha^{\frac{22}{39}} \
\dot{M}^{-\frac{10}{39}} \
M^{\frac{1}{3}} 
\\

\\
\ \ \ \ \  \ =
1.4 \times 10^{17}\alpha ^{\frac{22}{39}} \dot M_{-1}^{-\frac{10}{39}} 
M_8^{\frac{1}{13}}
\end{array}
\end{equation}
\bigskip
\newpage 
\begin{center}
{\bf Region  {\it a} for $\alpha$--discs }
\end{center}
\begin{equation}
\begin{array}{l}
R_{\rm sg}= 4.6733 \times 10^{-9} 
\kappa_{\rm es}^{\frac{2}{3}}
\alpha^{\frac{2}{9}}
\dot{M}^{\frac{4}{9}}\
M^{\frac{1}{3}} 
\\

\\
\ \ \ \ \  \ =
2.6 \times 10^{16}\alpha ^{\frac{2}{9}} \dot M_{-1}^{\frac{4}{9}} 
M_8^{\frac{7}{9}}
\end{array}
\end{equation}
\bigskip
\begin{center}
{\bf Region  {\it a} for $\beta$--discs }
\end{center}
\begin{equation}
\begin{array}{l}
R_{\rm sg}= 7.6518 \times 10^{-2} 
\kappa_{\rm es}^{\frac{1}{2}}
\alpha ^{\frac{1}{3}}
\dot{M}^{\frac{1}{6}}\
M^{\frac{1}{3}} 
\\

\\
\ \ \ \ \  \ =
4.5 \times 10^{16}\alpha ^{\frac{1}{3}} \dot M_{-1}^{\frac{1}{6}} 
M_8^{\frac{1}{2}}
\end{array}
\end{equation}
\bigskip
\begin{center}
{\bf  Region  {\it c}$^+$}
\end{center}
\begin{equation}
\begin{array}{l}
R_{\rm sg}= 1.3615 \times 10^{20} 
\kappa_{\rm o}^{\frac{1}{6}}
\alpha^{\frac{13}{18}}
\dot{M}^{-\frac{5}{9}} \
M^{\frac{1}{6}} 
\\

\\
\ \ \ \ \  \ =
1.1 \times 10^{17}\alpha ^{\frac{13}{18}} \dot M_{-1}^{-\frac{5}{9}} 
M_8^{-\frac{7}{18}}
\end{array}
\end{equation}
\bigskip
\begin{center}
{\bf  Region  {\it b}$^+$}
\end{center}
\begin{equation}
\begin{array}{l}
R_{\rm sg}= 1.3118 \times 10^{25}
\kappa_{\rm es}^{\frac{1}{3}}
\alpha^{\frac{2}{3}}
\dot{M}^{-\frac{1}{3}} 
\\

\\
\  \ \ \ \ \ = 
3.1 \times 10^{16}\alpha ^{\frac{2}{3}} \dot M_{-1}^{-\frac{1}{3}}
M_8^{-\frac{1}{3}}
\end{array}
\end{equation}
\bigskip
\newpage 
\begin{center}
{\bf  Region  {\it b}$^{*+}$ (for $\alpha$--discs) }
\end{center}
\begin{equation}
\begin{array}{l}
R_{\rm sg}  =  8.0208\times 10^{14}
\kappa_{\rm o}^{\frac{8}{3}} \
\alpha^{\frac{14}{9}}
\dot{M}^{\frac{10}{9}} \
M^{-\frac{7}{3}}
\\

\\
\ \ \ \ \  \ =
2.1 \times 10^{13}\alpha ^{\frac{14}{9}} \dot M_{-1}^{\frac{10}{9}}
M_8^{-\frac{11}{9}}
\end{array}
\end{equation}
\bigskip
\begin{center}
{\bf  Region  {\it b}$^{*+}$ (for $\beta $--discs) }
\end{center}
\begin{equation}
\begin{array}{l}
R_{\rm sg}  =7.2272 \times 10^{18}  
\kappa_{\rm o}^{\frac{4}{5}} \
\alpha^{\frac{14}{15}}
\dot{M}^{-\frac{2}{15}} \
M^{-\frac{7}{15}} 
\\

\\
\ \ \ \ \  \ =
1.4 \times 10^{16}\alpha ^{\frac{14}{15}} \dot M_{-1}^{-\frac{2}{15}}
M_8^{-\frac{3}{5}}
\end{array}
\end{equation}

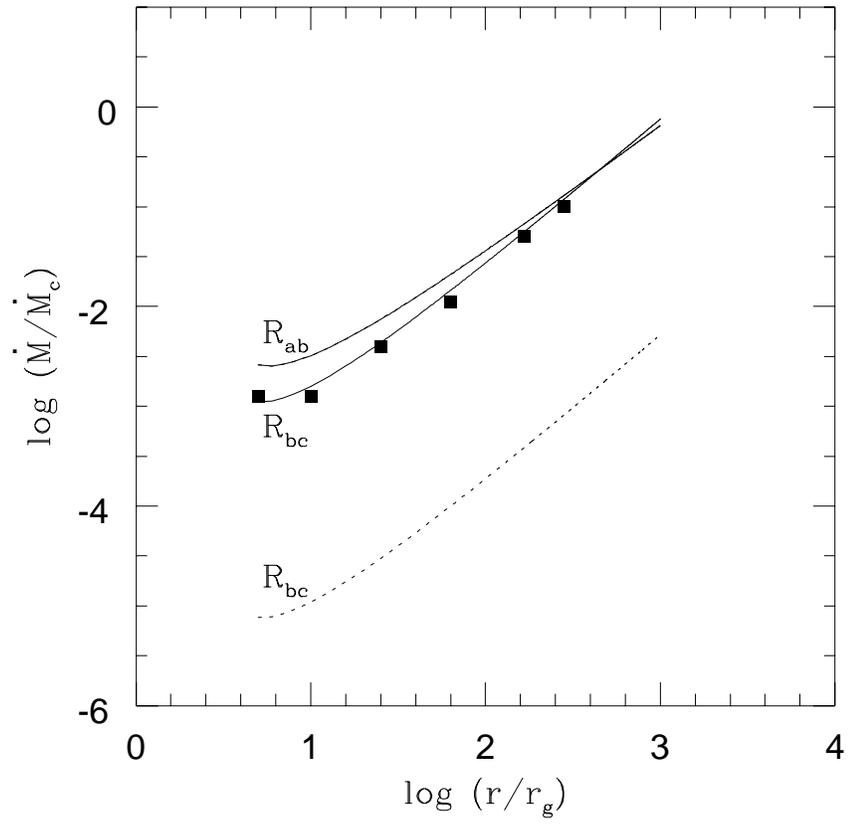
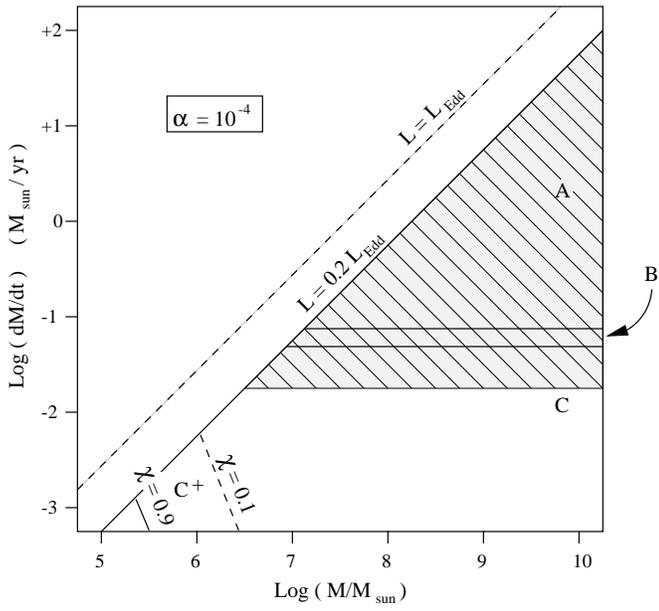
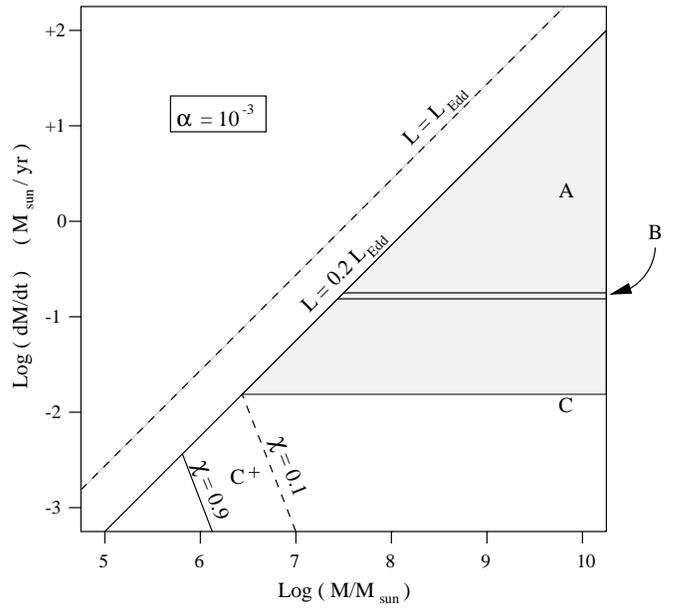
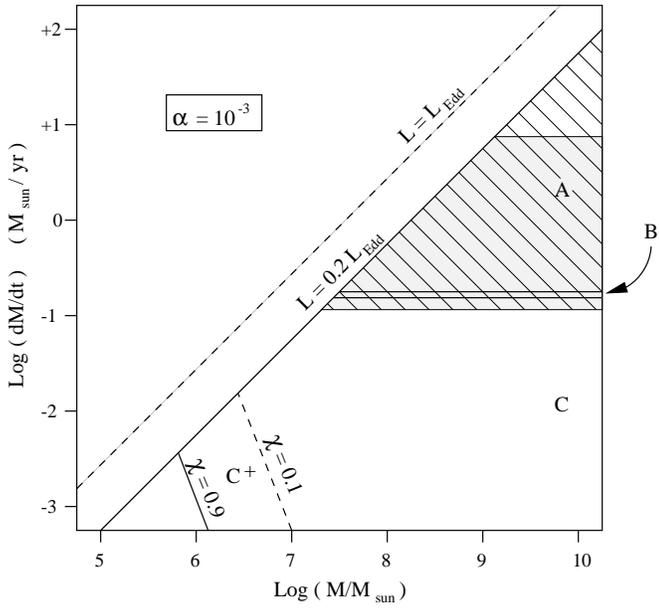
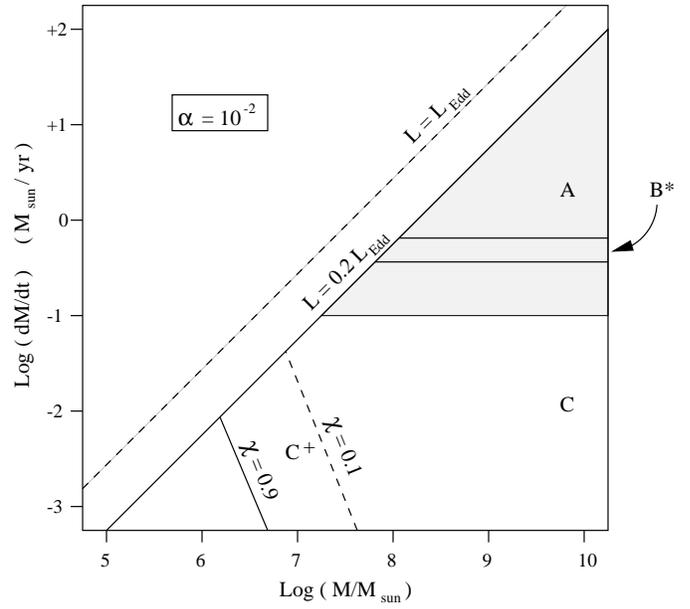
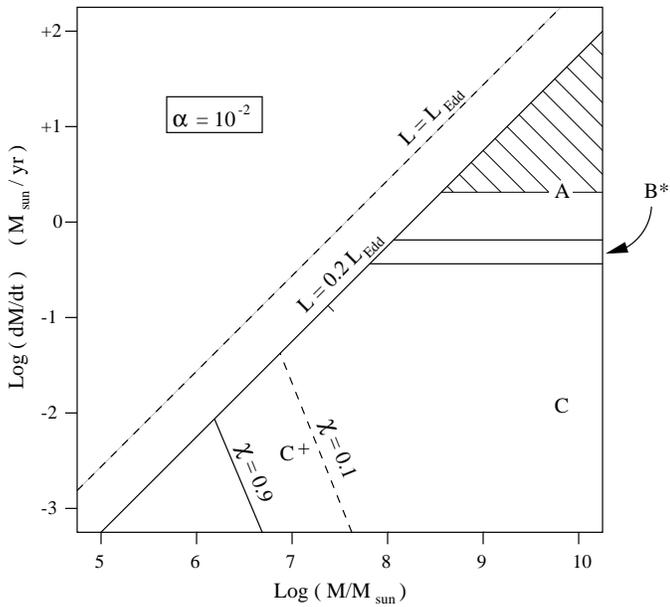
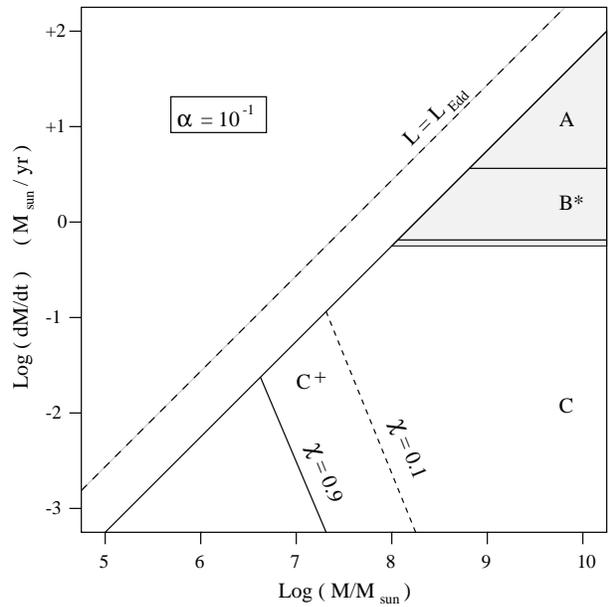
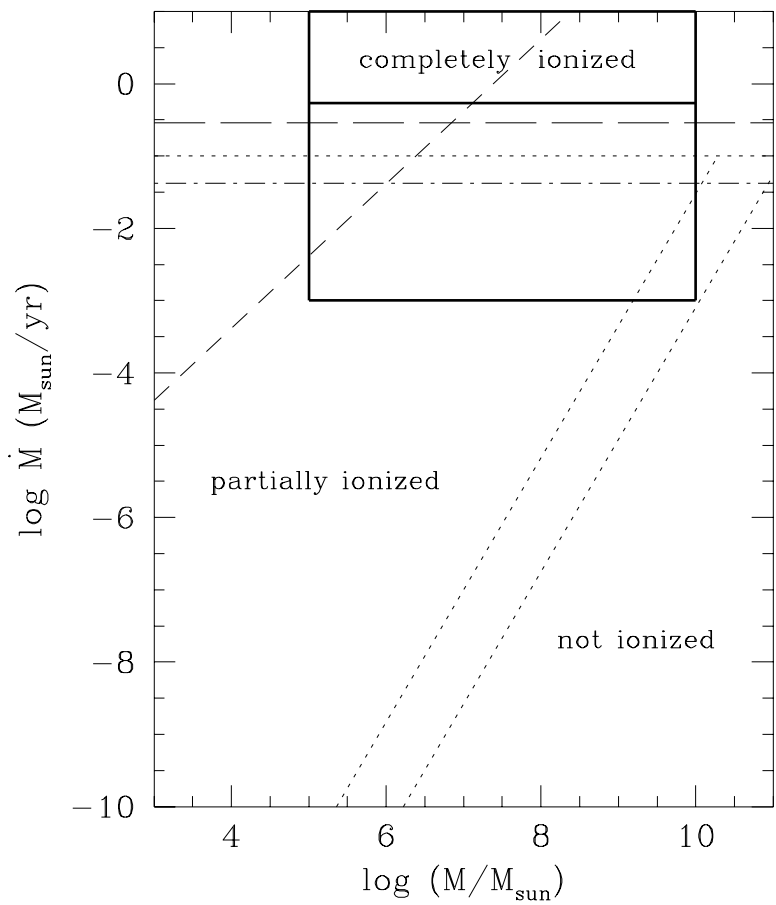
\end{document}